A material system for reliable low voltage anodic electrowetting

M. Khodayari, J. Carballo, Nathan B. Crane

Mechanical Engineering Department, University of South Florida

**ABSTRACT**

Electrowetting on dielectric is demonstrated with a thin spin-coated fluoropolymer over an aluminum electrode. Previous efforts to use thin spin-coated dielectric layers for electrowetting have shown limited success due to defects in the layers. However, when used with a citric acid electrolyte and anodic voltages, repeatable droplet actuation is achieved for 5000 cycles with an actuation of just 10 V. This offers the potential for low voltage electrowetting systems that can be manufactured with a simple low-cost process.

**Introduction**

Droplet manipulation has found applications in various fields including lab-on-chip [1-2], displays [3], optics [4], chip cooling [5], and energy harvesting [6]. Droplet shape and motion can be controlled by many methods including thermocapillary[7], dielectrophoresis[8] and magnetic forces [9], but electrowetting on dielectric (EWOD) [10-11] is a particularly promising method for low-cost, flexible and swift droplet actuation. In EWOD, droplet shape and equilibrium positions are changed due to an electromechanical effects of an electrical field applied across a fluid interface [12]. Typically, the electric field is created by applying a potential difference between an auxiliary electrode (often an electrode placed inside or on top of the droplet) and an electrode underneath a thin dielectric. Below a limiting saturation voltage, the wetting angle is given as[13]:

$$\cos\theta_1 = \cos\theta_0 + \varepsilon_0\varepsilon_r V^2 / 2\delta\gamma_{LO}. \qquad (1)$$

where $\theta_0$ and $\theta_1$ are the initial and electrowetting droplet angles, $V$ is the applied voltage, $\gamma_{LO}$ is the surface energy between the droplet and the second phase (here, hexadecane), $\delta$ is the dielectric thickness, and $\varepsilon_0\varepsilon_r$ the permittivity.

One challenge in EWOD is to manipulate the droplet with a low voltage, since many EWOD devices require tens to hundreds of volts while most electronics operate at much lower voltages. Those systems that have been demonstrated to operate below 30 V [14-17], typically require slow and/or expensive deposition processes such as atomic layer deposition [16], and chemical vapor deposition [16, 18] to create the dielectric layer. While spin-coated fluoropolymer dielectrics have been demonstrated, dielectric lifetime is often limited or not reported [15, 17, 19]. Low voltage operation is typically limited by declining dielectric properties in thin layers which could be the result of high electric field, local defects, and electrode oxidation[15]. However, recently, it has been shown that, in passivating metals, the electrodes can passivate at defects [18, 20-21]. These tests were done on Parylene coated on pre-oxidized aluminum layer (V > 16 volts) and thick Cytop dielectric layers over a passivating metal electrode (V > 60 volts). Electrolyte solutions (citric acid, tartaric acid) were chosen that form a passive oxidation layer with predominately anodic actuation voltages.

This paper reports on a reliable EWOD process with large contact angle modulation at voltages as low as 10 V using only a thin spin-coated polymer dielectric on bare aluminum. This compares favorably with previous thin spin-coated dielectrics requiring 15 V [14]for actuation. Other tests with thin flouropolymer dielectrics have shown very poor reliability at thin levels [15]. The authors are unaware of any previous reports of low voltage EWOD with demonstrated reliability and low voltage modulation.

**EXPERIMENTS**

A 300 nm thick aluminum layer was deposited by e-beam evaporation on a thermally oxidized silicon wafer and then various thickness of Cytop fluoropolymer (20 nm, 50 nm, 1100 nm) were deposited on the aluminum layer via spin-coating. The wafer was placed in hexadecane as the second phase and an 8 μl droplet of the electrolyte was pipetted on the wafer. Tested electrolytes include 0.1 M sodium sulfate ($Na_2SO_4$), 0.1 M tartaric acid and 0.1 M citric acid. A platinum auxiliary electrode was placed in the droplet. Then, using a Keithley 2612A SourceMeter, potential difference between the droplet and the wafer was applied in steps while the auxiliary electrode was grounded. The applied voltages were +13 V, +22 V and +60 V on wafers with 20, 50 and 1100 nm Cytop thicknesses, respectively. These voltages were chosen to achieve a 75 deg angle change. Horizontal images of the droplets were recorded at each voltage step. The contact angles were extracted from the images using the Drop Analysis plug-in for ImageJ [22]. Figure 1 shows the experimental setup schematics. Reliability was assessed by measuring the change in contact angle with repeated voltage application.

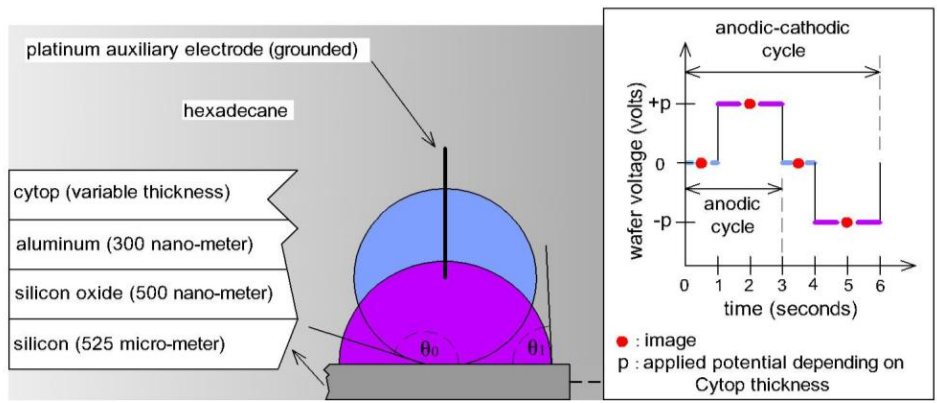

**Figure 1 Schematic of the experimental setup. Anodic-Cathodic electrowetting used alternating positive and negative voltage pulses as illustrated. In anodic electrowetting only positive pulses were used. These tests were performed on multiple wafers and reproducible results were observed.**

**DISCUSSION**

Aluminum films show diode-like current-voltage response [20-21] with high resistivity at positive potentials due to the passivation of the aluminum surface. If the proper voltage polarity is maintained, the leakage current is minimized and stable operation is possible. The role of voltage polarity in EWOD life is seen by comparing angle modulations from pure anodic cycles to alternating anodic and cathodic cycles as seen in Figure 2. EWOD with alternating voltage polarities (Figure 2, dashed lines) degrades rapidly with no angle response after a few cycles, presumably because of the high rate of cathodic reactions causing degradation of the aluminum surface and dielectric. In contrast, positive voltage alone exhibited only modest degradation of the zero voltage angle, and maintained steady operation for 1100 cycles of 2 sec voltage application (FIG. 3b, c).

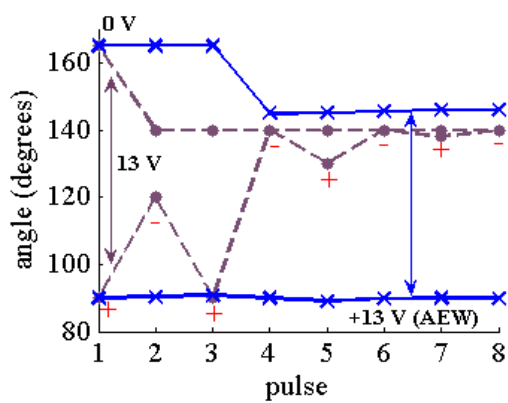

**Figure 2. (Color Online) Droplet electrowetting modulation with only +13 V pulses (solid line) and both +13 and -13 V pulses (dashed line) with 0.1 M citric acid. The systems with negative applied voltages (dashed lines) show rapid electrowetting degradation due to destructive effects of cathodic reactions.**

Typically, the EWOD reliability decreases with decreasing layer thickness. Figure 3(a) shows how a sodium sulfate shows much quicker degradation for thinner Cytop layers even when the voltage is adjusted to maintain a nearly constant contact angle. Similar results are seen with sodium chloride solutions. When a passivating electrolyte (citric acid) is used, the same dielectric shows larger angle changes for the same applied voltage and more stable operation (Figure 3(b)). Citric acid is known to form an effective passivating oxide on aluminum [23-24]. Other solutions such as tartaric acid [23, 25-26] that form passivating oxides also show good performance though tartaric acid electrolytes did not have as long of life as citric acid.

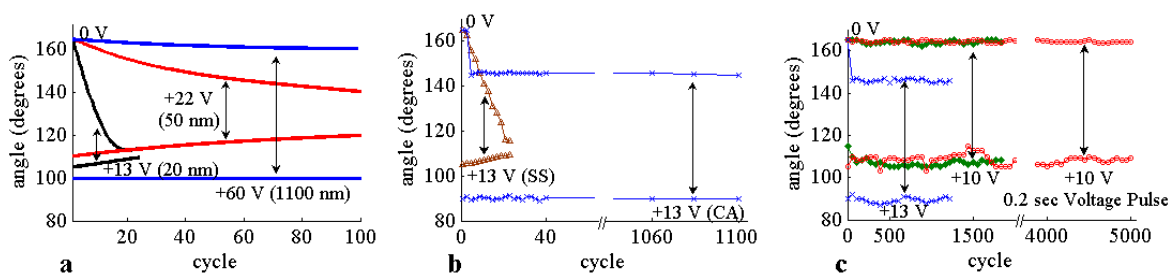

**Figure 3. (Color Online) Contact angle with and without applied voltages graphed for repeated cycles. In these graphs CA and SS indicate 0.1 M citric acid and 0.1 M sodium sulfate solution. a) Contact angle change for repeated cycles with different Cytop thickness with $Na_2SO_4$ electrolyte. In this figure fitted curves have been shown for better clarity. The tests were performed on 1100, 50 and 20 nm Cytop thicknesses with electrowetting voltages of +60, +22 and +13 volts, respectively. These voltages were chosen to achieve an electrowetting angle of 90 degrees with the citric acid solution, b) the effects of electrolyte composition on low voltage AEW lifetime on 20 nm Cytop and c) the effects of modulation voltage and voltage pulse length on low voltage AEW on 20 nm Cytop.**

Applied potential magnitude could cause some differences in anodic electrowetting (AEW). Figure 2 and Figure 3 b shows a drop in the contact angle at zero potential after just a few steps of the 13 V AEW. This could be due to charge entrapment [27]. However, when the voltage is slightly reduced from 13 V to 10 V, negligible angle change is seen over 1800 cycles (Figure 3c, Figure 4). After 1100$^{th}$ and 1650$^{th}$ cycles in respectively 13 and 10 V AEW, the droplets tended to jump off which is possibly related to a gradual aluminum oxidation and/or charge entrapment that would create differential dielectric properties over the wetting area. However, after the droplet motion, there was no visible damage to the dielectric or the electrode. In addition to potential magnitude, pulse duration can alter AEW life. When the same substrates were tested with shorter voltage pulses the droplet motion occurred at a higher number of cycles. For instance, with a voltage pulse length of 0.2 seconds, one tenth the time used in the other plotted data, droplet motion was not observed until after 5000 cycles (Figure 3c). Thus the length of time the voltage is applied affects the ultimate system life.

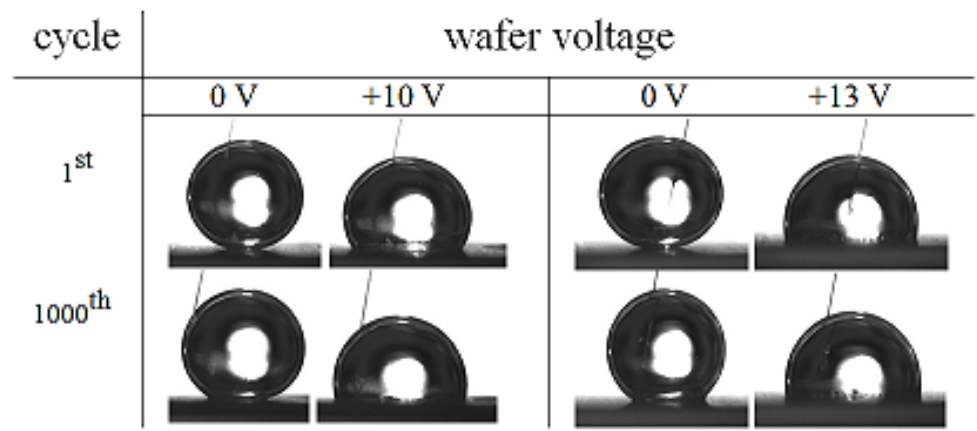

**Figure 4. Droplet images before and after AEW at the first and 1000$^{th}$ cycles (the droplet solution is 0.1 M citric acid).**

## CONCLUSION

Under anodic actuation voltages, valve-metal electrodes can be used with thin hydrophobic polymers as dielectric to achieve stable operation at low voltages. Typically, these layers have defects that permit significant electrochemical damage to the electrodes. However, combinations of electrodes and electrolyte known to form a passive oxide perform reliably even for thin spin-coated dielectric layers subject to many defects as long as the electrode is at a more positive potential than the droplet. A passivating oxide is believed to form at the defects to prevent continued electrode damage or droplet/electrode shorting.

This work used aluminum electrodes and compared the system performance for different droplet compositions. The success of this approach requires the proper selection of electrode and droplet composition to create a passivating oxide film. Metals such as Al, Hf, Nb, Ta, Ti, and Zr (sometimes referred to as valve metals) [28] can form such passivating oxide layers. Similar results should be possible with these metals. These electrode/electrolyte material combinations permit reliable electrowetting using simple dielectric deposition methods such as spin-coating and dip-coating processes. This enables experimental study of EW with simplified fabrication and could dramatically reduce the costs of producing commercial devices compared to alternative methods that utilize techniques like chemical vapor deposition or atomic layer deposition to deposit the dielectric.

Support for this work was provided through NSF Grant CMMI-092637, the state of Florida through the Florida Energy Systems Consortium (FESC), and by NACE International. The assistance of Dr. Alex Volinsky with manuscript preparation is gratefully acknowledged.

# REFERENCES

[1] Fan S-K, Yang H, Hsu W. Droplet-on-a-wristband: Chip-to-chip digital microfluidic interfaces between replaceable and flexible electrowetting modules. Lab on a Chip. 2011;11:343-7.

[2] Witters D, Vergauwe N, Vermeir S, Ceyssens F, Liekens S, Puers R, et al. Biofunctionalization of electrowetting-on-dielectric digital microfluidic chips for miniaturized cell-based applications. Lab on a Chip. 2011;11:2790-4.

[3] Heikenfeld J, Smith N, Dhindsa M, Zhou K, Kilaru M, Hou L, et al. Recent Progress in Arrayed Electrowetting Optics. Opt Photon News. 2009;20:20-6.

[4] Smith NR, Hou L, Zhang J, Heikenfeld J. Fabrication and Demonstration of Electrowetting Liquid Lens Arrays. J Display Technol. 2009;5:411-3.

[5] Cheng JT, Chen CL. Active thermal management of on-chip hot spots using EWOD-driven droplet microfluidics. Experiments in Fluids. 2010;49:1349-57.

[6] Krupenkin T, Taylor JA. Reverse electrowetting as a new approach to high-power energy harvesting. Nat Commun. 2011;2:448.

[7] Pratap V, Moumen N, Subramanian RS. Thermocapillary Motion of a Liquid Drop on a Horizontal Solid Surface. Langmuir. 2008;24:5185-93.

[8] Tianzhun W, Suzuki Y, Kasagi N, Kashiwagi K. Oil droplet manipulation using liquid dielectrophoresis on electret with superlyophobic surfaces.  Micro Electro Mechanical Systems (MEMS), 2010 IEEE 23rd International Conference on2010. p. 1055-8.

[9] Long Z, Shetty AM, Solomon MJ, Larson RG. Fundamentals of magnet-actuated droplet manipulation on an open hydrophobic surface. Lab on a Chip. 2009;9:1567-75.

[10] Ren H, Fair RB, Pollack MG. Automated on-chip droplet dispensing with volume control by electro-wetting actuation and capacitance metering. Sensors and Actuators B: Chemical. 2004;98:319-27.

[11] Sung Kwon C, Hyejin M, Chang-Jin K. Creating, transporting, cutting, and merging liquid droplets by electrowetting-based actuation for digital microfluidic circuits. Microelectromechanical Systems, Journal of. 2003;12:70-80.

[12] Jones TB. An electromechanical interpretation of electrowetting. Journal of Micromechanics and Microengineering. 2005;15:1184.

[13] Mugele F, Baret JC. Electrowetting: from basics to applications. Journal of Physics: Condensed Matter. 2005;17:705-74.
9